\documentclass[twocolumn,showpacs,preprintnumbers,amsmath,amssymb]{revtex4}
\usepackage{tabularx,graphicx}\begin{document}
\newcommand{\beq}{\begin{equation}}
\newcommand{\eeq}{\end{equation}}
\newcommand{\beqn}{\begin{eqnarray}}
\newcommand{\eeqn}{\end{eqnarray}}
\newcommand{\bmath}{\begin{subequations}}
\newcommand{\emath}{\end{subequations}}
\newcommand{\bra}[1]{\langle #1|}
\newcommand{\ket}[1]{|#1\rangle}

\title{The origin of the Meissner effect in new and old superconductors}
\author{J. E. Hirsch }
\address{Department of Physics, University of California, San Diego,
La Jolla, CA 92093-0319}

\begin{abstract} 
It is generally believed that superconducting materials are divided into two classes: `conventional' and `unconventional'. Conventional superconductors (the elements and thousands of compounds including $MgB_2$) are described by conventional London-BCS-Eliashberg electron-phonon theory. There is no general agreement as to what mechanism or mechanisms describe `unconventional' superconductors such as the heavy fermions, organics, cuprate and pnictide families. However all superconductors, whether
`conventional' or `unconventional', exhibit the Meissner effect.
I  argue that there is a single mechanism of superconductivity for all materials, that  explains the Meissner effect and differs from the conventional mechanism in several fundamental aspects: it says that superconductivity is driven by lowering of kinetic rather than potential energy of the charge carriers,  it requires conduction by holes rather than electrons in the normal state,
and it predicts a non-homogeneous rigid charge distribution and an electric field in the interior of superconductors. Furthermore I argue that neither  the conventional mechanism nor any of the other proposed unconventional mechanisms can explain
the Meissner effect. Superconductivity in  materials is discussed in the light of these concepts,
some experimental predictions, connections to Dirac's theory, and  connections to the superfluidity of $^4He$.    \end{abstract}
\maketitle 

\section{introduction}

BCS theory still reigns as the undisputed explanation of superconductivity in `conventional' superconductors\cite{bcs50}.
In a somewhat circular argument, a `conventional' superconductor is $defined$ to be a superconductor described by
BCS theory.  In addition there are by now at least $10$ different classes of materials that are generally believed to be
`unconventional', i.e. not described by BCS theory\cite{validity,norman}. Yet the belief that conventional BCS electron-phonon theory describes
the simplest materials, elements and compounds, remains unwavering\cite{cohen}, irrespective of the fact that in relative terms
the phase space of `conventional' superconductors is rapidly  shrinking. It should also be stressed that
the supposedly `unconventional' superconductors are not necessarily characterized by having a large critical 
temperature, since they are found e.g. among heavy fermion materials with $T_c$'s of a few degrees,
organic superconductors of the order of $10$ degrees, electron-doped cuprates and many iron-pnictide materials
with $T_c$'s below $30$ degrees, all lower than the supposedly conventional superconductor $MgB_2$ with
$T_c=39K$. And certainly there is no single `unconventional mechanism' proposed to describe all unconventional 
superconductors: new mechanisms are being proposed that apply specifically to one family only, e.g. the cuprates,
or the iron pnictides, or the heavy fermion superconductors.

However all superconductors, whether conventional or not, exhibit the Meissner effect. I  argue that BCS theory cannot explain
the Meissner effect\cite{missing}, so it cannot explain $any$ superconductor. Furthermore, none of the unconventional mechanisms
proposed to explain `unconventional' superconductivity has addressed the question of how to explain the Meissner
effect. I argue that none of these mechanisms describe any superconductor because they cannot explain the
Meissner effect.

I propose that the Meissner effect can only be explained if: (i) superconductivity is driven by lowering of the
kinetic energy of the charge carriers\cite{color}, and (ii) superconductors expel negative charge from the interior to the surface
in the transition to superconductivity\cite{expulsion0}. This physics results in a macroscopically inhomogeneous charge
distribution\cite{expulsion}   and in the existence of macroscopic zero-point motion which manifests itself in the
form of a spin current\cite{spincurrent} in the ground state of superconductors. Neither BCS theory nor London electrodynamic theory
describe this physics. Nevertheless, parts of both BCS theory and London theory are  undoubtedly  correct.

The points (i) and (ii) are intimately connected. Kinetic energy lowering means, e.g. via Heisenberg's uncertaintly
principle, $expansion$ of the electronic wave function which in turn $implies$ outward motion of
negative charge. That outward motion of negative charge explains the generation of the Meissner current is
immediately seen from the action of the Lorentz force\cite{lorentz}. 
That the Meissner effect is impossible in the absence of outward motion of charge is immediately seen from the
equations of motion\cite{japan1} and from the fact that there is no other source of electromotive force\cite{emf}. That kinetic energy lowering drives superconductivity
follows from the fact that the Meissner effect cannot occur unless there is outward motion of negative charge;
outward motion of negative charge implies charge separation, hence increase in potential energy, 
so the `emf' driving it\cite{emf} has to be lowering of kinetic energy.

Furthermore our explanation of superconductivity is supported by 
 the   close relationship that is known to exist between superconductors and superfluids, e.g.
$^4He$\cite{londonbooks}. Both are macroscopic quantum phenomena\cite{londonbooks}. Both exhibit frictionless
flow with vanishing generalized vorticity.
Kinetic energy lowering drives the superfluid transition in $^4He$\cite{helium}, so it is natural to conclude
that it also drives the superconducting transition in superconductors.

We quote from the preface of London's book on superfluids, Vol. II\cite{londonbooks}: {\it ``That something strange happens to liquid helium at
about $2.2^oK$ was noticed by Kammerlingh Onnes as early as 1911. He found that when the liquid is cooled below
that temperature it starts expanding instead of continuing to contract, thus deviating from the behavior of most substances''.}
Indeed, the expansion of $^4He$ below the critical temperature is clear indication that the transition is
driven by kinetic energy lowering, and it parallels\cite{helium} the wavefunction expansion and charge expulsion that we  
propose exists in   superconductors, also driven by kinetic energy lowering.

Instead, BCS theory and conventional London electrodynamic theory imply that superconductivity is 
 driven by qualitatively different physics, that is non-existent in liquid $^4He$. Nobody has ever proposed that 
what drives conventional superconductivity within BCS theory, ``a footling small
interaction between electrons and lattice vibrations''\cite{mend}, has anything to do with the superfluidity of $^4He$. 
And if charge moves outward driven by kinetic energy lowering, conventional London electrodynamics will not apply because
it requires  absence of electric fields inside superconductors.

Remarkably, the only suggestion we could find in the scientific literature before the high $T_c$ era  that kinetic energy lowering has 
anything to do with superconductivity is in the preface of London's book on superconductivity\cite{londonbooks},
where he writes: {\it ``ÒIt is not necessarily a configuration
close to the minimum of the potential energy
(lattice order) which is the most advantageous one for
the energy balance, since by virtue of the uncertainty
relation the kinetic energy also comes into play. If the
resultant forces are sufficiently weak and act between
sufficiently light particles, then the structure possessing
the smallest total energy would be characterized by
a good economy of the kinetic energy''}.
However, in the remainder of the book no mention whatsoever is made of how ``a good economy of the 
kinetic energy''  would play any role in superconductivity. 

The reason that London said this in his preface is presumably that he 
had superfluid $He$ in mind together with the close relation in $other$ properties of 
superconductors and superfluids. There was however
no indication at that time either from experiment or from theory that kinetic energy lowering had anything to do with the transition to
superconductivity.

That has changed by now. Since the early days of the theory of hole superconductivity discussed in this paper it was clear that electron-hole asymmetry
was intimately related to kinetic energy: the pairing interaction (correlated hopping) was termed $\Delta t$\cite{corrhop}, 
indicating its relation with the hopping amplitude $t$ and kinetic energy (in contrast to other works\cite{x,x2,k}).
The relation between kinetic energy lowering, wavefunction expansion and charge asymmetry became clearer when
it was realized that superconductors expel $negative$ charge from their interior to the surface\cite{expulsion0,expulsion}. Experimentally, evidence 
for kinetic energy lowering was found in optical properties of  cuprates and pnictides\cite{opticalexp} as predicted theoretically
several years before the experiments\cite{apparent}.

\section{electromotive force}

An electromotive force is a non-electric force that moves electric charges $against$ the direction dictated by electric fields.
In a voltaic cell, an electromotive force moves positive charges from the negative to the positive electrode (raising their electric
potential energy). Similarly in the Meissner effect an electromotive force is needed to accelerate the electric charges near the surface carrying the 
developing Meissner current in
direction opposite to that dictated by the electric force generated by Faraday's law as the magnetic field lines are moving out\cite{lenz}.

Neither conventional BCS-London theory nor any of the unconventional theories of superconductivity proposed in recent years (except for the one discussed here)
offer an explanation of what this electromotive force is in superconductors. In other words, while they describe the initial and final state, they cannot describe the
$process$ by which the system evolves from the initial to the final state. In the absence of such an electromotive force,  metals in the presence of a magnetic field
would never become superconducting. But they do.

There is a limited number of possibilities offered by the known laws of physics. We know of four fundamental forces:
gravitation, strong, weak and electromagnetic. It is almost obvious that neither gravitational, nor strong (nuclear) nor weak interactions can play a role.
We are left with electromagnetic, however we are seeking a non-electromagnetic electromotive  force.

There is in fact a known fifth force in nature: the ``quantum force''. A quantum particle confined to a finite volume exerts``quantum pressure" against the confining walls.
This quantum pressure, times the area over which it acts, gives us a force.
We have argued in reference \cite{emf} that this is the electromotive force that explains both the physics of voltaic cells and the Meissner effect.

This then leaves us with just two possibilities: either there is another force in physics, as yet unknown, that explains the electromotive force manifested
in the Meissner effect. If so, the proponents of the conventional theory of superconductivity or of other unconventional theories should explain what that force is.
Or, the quantum force mentioned above explains the Meissner effect. The theory described in this paper proposes the latter, and explores the consequences
of this for the physics of superconductivity. We find that many resulting properties of superconductors are qualitatively different from the predicted properties
of superconductors within conventional BCS-London theory. If proponents of the conventional theory were to argue that this quantum force also explains
the Meissner effect in the conventional framework, they have to explain how one would avoid the other consequences of this physics that our theory
predicts to be unavoidable.

\section{kinetic energy lowering, radial expansion and the meissner effect}

  \begin{figure}
 \resizebox{8.5cm}{!}{\includegraphics[width=9cm]{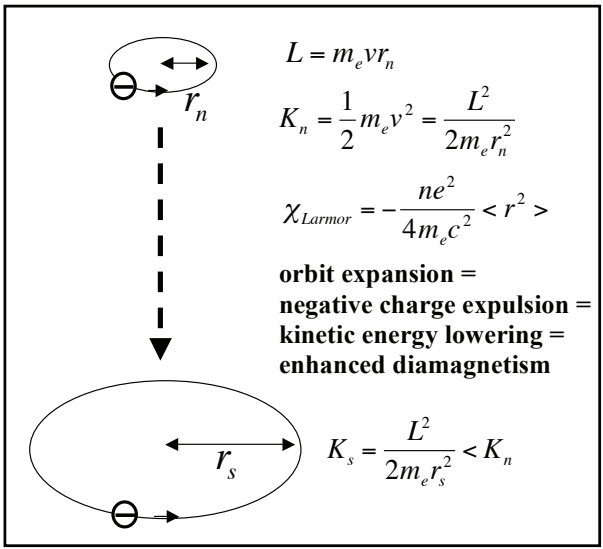}}
 \caption {The essence of superconductivity. An electron in an expanding orbit with fixed angular momentum
 lowers its kinetic energy ($K_s<K_n$), increases its diamagnetic susceptibility and causes expulsion of negative charge.
 The top orbit represents the normal state, with $r_n=k_F^{-1}$, the bottom one the 
 superconducting state, with $r_s=2\lambda_L$.}
 \label{figure2}
 \end{figure}

Figure 1 shows what we propose  is 
the essential physics of the Meissner effect and hence of superconductivity.  As the electronic orbit expands from radius $r_n$ to $r_s$, with fixed angular momentum $L$,
the kinetic energy decreases, diamagnetism increases, and negative charge moves outward. From  the Larmor expression for
the diamagnetic susceptibility
\beq
\chi_{Larmor}=-\frac{ne^2}{4m_e c^2} <r^2>
\eeq
one obtains Landau diamagnetism for $r=k_F^{-1}$ ($k_F=$ Fermi momentum) and perfect diamagnetism for
$r=2\lambda_L$, with $\lambda_L$ the London penetration depth\cite{kinetic}. 
This is easily seen using the expressions for the electronic density of states $g(\epsilon_F)=3n/2\epsilon_F$ and the standard expression for the
London penetration depth 
\beq
\frac{1}{\lambda_L^2 }= \frac{4\pi ne^2}{m_e c^2} .
\eeq
Thus, the transition to superconductivity involves
expansion of electronic orbits from $r_n=k_F^{-1} \sim \AA$ to $r_s=2\lambda_L\sim 1000\AA$.
The correct theory of superconductivity should contain the physics depicted in Fig. 1. BCS theory does not.

The Hamiltonian for the conduction electrons in a metal is
\beq
H=K+U_{pot}
\eeq
The first term is the electronic kinetic energy, given by
\beq
K= \sum_i(-\frac{\hbar^2}{2m_e})\nabla_i^2
\eeq
and the second term is the potential energy,  given by the sum of electron-ion  and electron-electron interactions
\beq
 U_{pot} =U_{el-ion}+U_{el-el}   .
 \eeq
Conventional BCS theory attributes the energy lowering in going from the normal to the superconducting state
 to $U_{el-ion}$ together with the ion dynamics. Unconventional mechanisms proposed to describe
 new superconductors propose that the energy lowering originates in $U_{el-el}$, usually involving
 magnetic mechanisms. Instead, we propose that superconductivity in all materials originates in the fact that
 the average electronic kinetic energy is lower in  the superconducting state than in the normal state:
\beq
 \bra{ \Psi_{super} } K \ket{\Psi_{super}} <  \bra{ \Psi_{normal} } K \ket{\Psi_{normal}} 
\eeq
while the average potential energy is higher
\beq
 \bra{ \Psi_{super} } U_{pot}  \ket{\Psi_{super}}  >  \bra{ \Psi_{normal} }   U_{pot} \ket{\Psi_{normal}} 
 \eeq
 albeit by a lesser amount, so that the difference yields the condensation energy of the superconductor.
 
 We can  understand the energetics involved and its connection with orbit expansion by looking
 at a two-electron atom, i.e. $H^-$ or $He$, with atomic number $Z=1$, $Z=2$ respectively. Assuming the variational wavefunction
 \beq
\Psi_{r_0} (\bold{r}_1,\bold{r}_2)=\varphi_{r_0}  (r_1)\varphi _{r_0} (r_2)  
\eeq
with
\beq
\varphi _{r_0} (r)=(\frac{1}{r_0^3 \pi})^{1/2}e^{-r/r_0}
\eeq
we have
\bmath
\beq
\bra{\Psi_{r_0} }K \ket{\Psi_{r_0}}=2\frac{\hbar^2}{2m_e r_0^2}
\eeq
\beq
\bra{\Psi_{r_0} }U_{el-ion}\ket{\Psi_{r_0}}= -2\frac{Ze^2}{r_0}
\eeq
\beq
\bra{\Psi_{r_0} }U_{el-el}\ket{\Psi_{r_0}}=\frac{5}{8} \frac{e^2}{r_0}
\eeq
\emath

Therefore, if in going from the normal to the superconducting state $r_0$ changes from $r_n$ to $r_s>r_n$ we have from
Eqs. (10)
\bmath
\beq
\bra{\Psi_{r_s} }K \ket{\Psi_{r_s}} < \bra{\Psi_{r_n} }K \ket{\Psi_{r_n}}
\eeq
\beq
\bra{\Psi_{r_s} }U_{pot} \ket{\Psi_{r_s}}   > \bra{\Psi_{r_n} }U_{pot} \ket{\Psi_{r_n}}
\eeq
\emath
(the latter valid as long as $Z>5/16$) so that the kinetic energy decreases and the potential energy increases. More specifically,
\bmath
\beq
\bra{\Psi_{r_s} }U_{el-ion} \ket{\Psi_{r_s}}  >  \bra{\Psi_{r_n} }U_{el-ion} \ket{\Psi_{r_n}}
\eeq
\beq
\bra{\Psi_{r_s} }U_{el-el} \ket{\Psi_{r_s}} < \bra{\Psi_{r_n} }U_{el-el} \ket{\Psi_{r_n}}
\eeq
\emath
so the potential energy increase results from increased electron-ion energy partially compensated by a decrease in
electron-electron energy.

The minimum in the total energy occurs for
\beq
\bar{r}_0=\frac{a_0}{\bar{Z}}
\eeq 
with $a_0=\hbar^2/(m_e e^2)$ the Bohr radius and 
\beq
\bar{Z}=Z-\frac{5}{16}
\eeq
Therefore, in order for the minimum value of the energy to correspond to the mesoscopic scale $2\lambda_L$, the 
`effective' nuclear charge  $\bar{Z}$ has to be  very small:
\beq
\bar{Z}\sim \frac{a_0}{2\lambda_L}             .
\eeq
This illustrates that orbit expansion and the Meissner effect will be associated with situations where carriers propagate
through {\it negatively charged anions}, where the `effective' nuclear charge $\bar{Z}$  is close to zero, such as 
$O^=$, $As^{---}$, $S^=$, $Se^=$ and $B^-$. It is not surprising therefore that high $T_c$ superconductivity
is found in cuprates, iron pnictides, iron chalcogenides and magnesium diboride.

When the orbits expand, they become highly overlapping, so they have to become phase coherent to avoid collisions that
would raise the potential energy even further. In contrast, the small orbits in the normal state are non-overlapping and hence
can have arbitrary phases. Therefore, the entropy is larger in the normal state (small orbits) than in the superconducting state
(large orbits) and for this reason the normal state is favored at high temperatures and the superconducting state at low
temperatures.

Finally, to achieve a low value of the `effective' nuclear charge $\bar{Z}$ requires the electronic conduction band to be almost
full so that the negative electrons cancel the positive charge of the ions. Thus superconductivity will occur if there are
almost full bands, giving rise to dominance of positive Hall coefficient (hole conduction) in the normal state.

\section{theory of hole superconductivity}
   \begin{figure}
 \resizebox{8.5cm}{!}{\includegraphics[width=9cm]{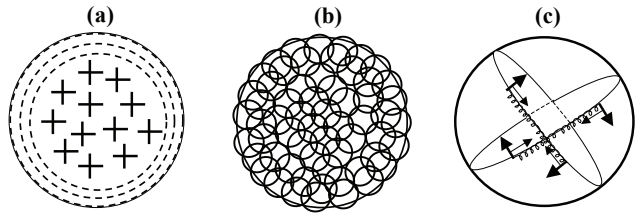}}
 \caption{Illustration of three key aspects of the physics of superconductors proposed here. (a) Superconductors expel negative charge from their
 interior to the region near the surface; (b) Carriers reside in mesoscopic overlapping orbits of radius $2\lambda_L$
 ($\lambda_L$=London penetration depth); (c) A spin current flows near the surface of superconductors (the arrow perpendicular to the orbit denotes the direction
 of the electron magnetic moment). }
 \label{figure2}
 \end{figure}
The theory of hole superconductivity\cite{holetheory} is consistent with the physics of the Meissner effect discussed in Sect. II and depicted in
Fig. 1. More specifically, it 
predicts the physics shown schematically in Fig. 2. Superconductors expel negative charge
from the interior to the surface, due to the expanding orbits, giving rise to an excess negative charge density\cite{electrospin}
\beq
\rho_-=en_s\frac{\hbar}{4m_e c \lambda_L}
\eeq
within a London penetration depth of the surface. An electric field exists in the interior pointing outward, with maximum value
given by
\beq
E_m=-\frac{\hbar c}{4e\lambda_L^2} .
\eeq

Electrons within a London penetration depth of  the surface carry a spin current, with velocity given by\cite{sm}
\beq
\vec{v}_\sigma^0=-\frac{\hbar}{4m_e\lambda_L} \vec{\sigma} \times\hat{n}
\eeq
with $\hat{n}$ the outward-pointing normal to the surface. 
It originates in the superposition  of    rotational zero-point motion of electrons in the $2\lambda_L$ orbits throughout the bulk of the system, that cancels out in the interior
but not near the surface, just like Amperian surface currents originate in the sum of local magnetic dipole currents in the bulk of a magnetized material.
The electrodynamic equations governing the behavior of
electric and magnetic fields, charge and spin currents, are given in ref. \cite{electrospin}.

The proposal that superconductors eject electrons from their interior to the surface
depicted in Fig. 2(a), whether or not magnetic fields are present, has not yet been experimentally verified. However we 
argue that it is clearly illustrated in the situation shown in Fig. 3. When current flows from a normal conductor into a superconductor, flow lines
go to the surface. Since the current is carried by charge carriers, charge carriers entering the superconductor in the interior region have to flow to
the surface. The ``London moment'' experiments\cite{londonmoment}  and the
``gyromagnetic effect'' experiments\cite{gyro} show that charge carriers in the superconducting state are always $electrons$\cite{londonmomentyo}, hence
it is electrons (with their $negative$ charge) that flow to the surface as they enter the superconducting region. At the same time, magnetic field lines, which are circles throughout the interior in 
the normal region, are pushed out to the surface. It is only a small additional step to conclude that electrons 
in superconductors move to the surface whether or not a charge current is flowing, carrying any existing interior magnetic field lines with them.  

   \begin{figure}
 \resizebox{8.5cm}{!}{\includegraphics[width=9cm]{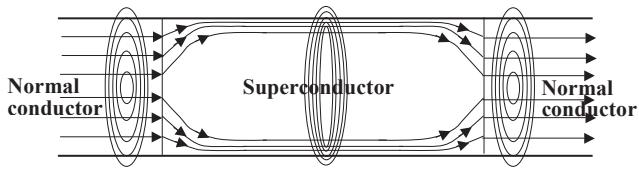}}
 \caption{Current distribution in a superconducting wire fed by normal conducting leads. As electrons enter the superconducting region their
 velocities acquire a radial component and  charge moves towards the surfaces together with magnetic field lines (circles). }
 \label{figure2}
 \end{figure}

The orbital angular momentum of electrons in orbits of radius $2\lambda_L$ and speed $v_\sigma^0$ is $L=m_ev_\sigma^0(2\lambda_L)=\hbar /2$. Thus, it reflects the intrinsic spin angular
momentum of the electron, which itself can be thought of as an orbital motion at speed $c$\cite{z3,z4}  in an orbit of radius 
$r_q=\hbar/2m_ec$, the ``quantum electron radius''\cite{qer}(as opposed to the classical electron radius $r_c=e^2/2m_ec$).  

There exists a remarkable parallel
in the physics at the three different length scales $r_q$, $a_0=\hbar^2/m_e e^2$ (Bohr radius) and $2\lambda_L$, corresponding to the length scale of the
electron, the atom and the superconductor. Slater has already remarked\cite{slater} that for superconductors
``the orbits must be of order of magnitude of $137$ atomic diameters''. If we asume that the carrier density is given by
\beq
n_s=\frac{1}{4\pi a_0^3}
\eeq
it yields a London penetration depth (using Eq. (2))
\beq
\lambda_L=\frac{a_0}{\alpha}=137a_0
\eeq
with $\alpha=e^2/\hbar c$ the fine structure constant. The atomic or band length scale $a_0$ is then precisely  the geometric mean of the electron length scale $r_q$ and the
superconducting length scale $2\lambda_L$. It is as if an expansion takes place

$r_q$ $\rightarrow$ $a_0$ $\rightarrow$ $2\lambda_L$

\noindent with scaling factor $2/\alpha$, i.e.
\beq
2 \lambda_L=\frac{2}{\alpha} a_0=(\frac{2}{\alpha}) (\frac{2}{\alpha}) r_q
\eeq
Similarly the corresponding energy scales are obtained by multiplying the Dirac energy scale $E_{Dirac}=2m_ec^2$ by powers of the fine structure constant
\beq
E_{sc}= (\frac{\alpha}{2} )^2 E_{band}= (\frac{\alpha}{2} )^2 (\frac{\alpha}{2} )^2E_{Dirac} 
\eeq
where the energies of quantum confinement over these length scales are given by
\bmath
\beq
E_{Dirac}=\frac{\hbar^2}{2m_er_q^2}=2m_e c^2=1.022MeV
\eeq
\beq
E_{band}=\frac{\hbar^2}{2m_ea_0^2}=\frac{e^2}{2a_0}=13.6eV
\eeq
\beq
E_{sc}=\frac{\hbar^2}{2m_e(2\lambda_L)^2}=2\nu=181\mu eV
\eeq
\emath
where
\beq
\nu=\frac{\hbar^2 q_0^2}{4m_e}
\eeq
(with $q_0=(2\lambda_L)^{-1}$) is the kinetic energy lowering per electron obtained within our theory in the transition to the superconducting state
(the kinetic energy of electrons in the spin current is $(1/2)m_e (v_\sigma^0)^2=\nu/2$\cite{nu}).
Thus, spatial expansion with factor $(2/\alpha)$ and corresponding energy reduction by $(\alpha/2)^2$ is seen to connect these three very different realms.

The expelled charge density $\rho_-$ is related to the total superfluid charge density $en_s$ by the same factor\cite{electrospin}:
\beq
\rho_{-} = \frac{v_\sigma^0}{c} en_s= \frac{r_q}{2\lambda_L}en_s=(\frac{\alpha}{2})^2 en_s
\eeq
as is the spin current speed to the speed of light
\beq
v_\sigma^0=\frac{r_q}{2\lambda_L}c=(\frac{\alpha}{2})^2 c
\eeq
Within Dirac theory, the ratio of the small to the large component of the electron wave function is $\sim v/c$ in the non-relativistic limit.
Thus Eqs. (25) and (26) indicate that the expelled charge density $\rho_-$ reflects the small component of the electron wave function.
Denoting by $|\varphi>$ and $|\chi>$ the large and small components of the electron wavefunction,
$\rho_-\sim e<\chi | \varphi>$.

 \section{electronic zitterbewegung}
 Within Dirac's theory of the electron, the `instantaneous' velocity of the electron is always $c$, the speed of light\cite{z1,z2}: the time derivative of the
 position operator in the Heisenberg representation is $dx_k/dt=c\alpha_k$,with $\alpha_k$ the Dirac $\alpha-$matrices, and $\alpha_k^2=1$. The motion of the
 electron with average speed $v$ thus has superposed a rapidly oscillating component at speed $c$ (termed `Zitterbewegung' by Schr\"{o}dinger\cite{z1}), that has been
 interpreted as a circular motion of radius $r_q=\hbar/2m_ec$ giving rise to the spin angular momentum $\hbar/2$ and the electron magnetic moment\cite{z3,z4}.
 
 The rotational zero-point  motion in $2\lambda_L$ orbits with orbital angular momentum $\hbar/2$ predicted by our theory can be seen as an 
 amplified version of this microscopic Zitterbewegung.
 It is remarkable that for each spin
 component the spin current in the absence of applied magnetic field is
 \beq
 j_\sigma=\frac{n_s}{2} v_\sigma^0=\frac{\rho_-}{2}c
 \eeq
 using Eq. (25), hence it can be interpreted as originating in
  the excess electrons of each spin propagating at speed $c$ in opposite directions. Moreover, when a magnetic field is applied the currents 
 change to
 \beq
 j_\sigma=\frac{n_s}{2}(v_\sigma^0-\frac{e\lambda_L}{m_e c}\vec{\sigma}\cdot \vec{B})
  \eeq
  as the spin parallel (antiparallel) to $\vec{B}$ slows down (speeds up). 
   At the same time,
 it is found\cite{electrospin} that the excess charge density changes by the same factor:
 \beq
 \rho_\sigma=\frac{n_s}{2c}(v_\sigma^0-\frac{e\lambda_L}{m_e c}\vec{\sigma}\cdot \vec{B})
 \eeq
 i.e. the excess charge density of spin parallel (antiparallel) to $\vec{B}$ that slows down (speeds up) decreases (increases) in magnitude. Thus,
 \beq
 j_\sigma=\rho_\sigma c
 \eeq
 holds for any applied magnetic field lower than the critical field. In other words, the predicted ground state currents both with and without applied magnetic field
 can be seen as resulting from the electrons giving rise to the excess charge density moving at the speed of light, just as in Schr\"{o}dinger's Zitterbewegung.

\section{energetics and the virial theorem}
It has been argued that the virial theorem implies that the kinetic energy of electrons is necessarily $increased$ in going from the normal to 
the superconducting state\cite{singh,chester}, independent of what is the mechanism for superconductivity. According to the virial theorem\cite{fock}, in a system where the only interactions are Coulomb interactions
\beq
<K>=-\frac{1}{2}<U_{pot}>
\eeq
where $<>$ denotes expectation value with a wavefunction that is an eigenstate of the Schr\"{o}dinger equation. The total energy is then
\beq
E=<K>+<U_{pot}>=-<K>
\eeq
In going from the normal state (or from an ensemble of normal states at finite temperature above  $T_c$) to the superconducting state, the total energy
necessarily has to decrease. Equation (32) then would seem to indicate that the kinetic energy $increases$, in contradiction with the discussion in our previous
sections. 

However, Eq. (31) only holds if the Schr\"{o}dinger equation is assumed to apply without relativistic corrections.  As discussed in the previous section, relativity plays a key role in the theory of
hole superconductivity.  We can estimate the magnitude of deviation in the energetics predicted by the nonrelativistic virial theorem by the following argument. 
Electrons can be thought of as being ``spread out'' over a distance $r_q=\hbar/2m_ec$, so the Coulomb potential form $e^2/r$ is not valid when $r$ becomes 
comparable or smaller than $r_q$. The average of the Coulomb potential over this region is 
\beq
\delta U=<\frac{e^2}{r}>_{r_q}  \sim \frac{e^2 r_q^2}{a_0^3}=362\mu eV
\eeq
for a wavefunction extending over a distance $a_0$ (Bohr radius). The magnitude of this term is certainly large enough to be relevant to superconducting condensation
energies.

The term just discussed corresponds to the ``Darwin term'' in the non-relativistic limit of the Dirac equation. An equally important correction comes from the spin-orbit
interaction
\beq
U_{s.o.}=-\frac{e\hbar}{4m_e^2 c^2} \vec{\sigma}\cdot (\vec{E}\times\vec{p})   .
\eeq
The electric field $\vec{E}$ generated by the positive compensating ionic charge density $|e|n_s$ at distance $2\lambda_L$ is
\beq
E=2\pi |e| n_s (2\lambda_L)
\eeq
and taking $p=m_ev_\sigma^0$ and using Eq. (2), Eq. (34) yields
\beq
U_{s.o.}=\frac{\hbar^2}{4m_e(2\lambda_L)^2}=\nu
\eeq
which is the same as Eq. (24). It is also closely related to the correction Eq. (33), since using (from Eq. (21)) that $r_q(2\lambda_L)=a_0^2$:
\beq
U_{s.o.} = \frac{e^2 r_q^2}{4a_0^3} .
\eeq

In summary, we conclude that the argument that the virial theorem forbids kinetic energy driven superconductivity is invalid if relativity plays a key role
in superconductivity. As discussed here and in earlier work, relativity plays a key role in the theory of hole superconductivity. This is also seen from
the fact that within this theory the screening of electrostatic fields takes place over distance $\lambda_L$\cite{expulsion}, that involves the speed of light, rather than
over the Thomas Fermi length (that does not involve the speed of light)  as in the conventional theory.

\section{superconductivity in materials}

   \begin{figure}
 \resizebox{8.5cm}{!}{\includegraphics[width=9cm]{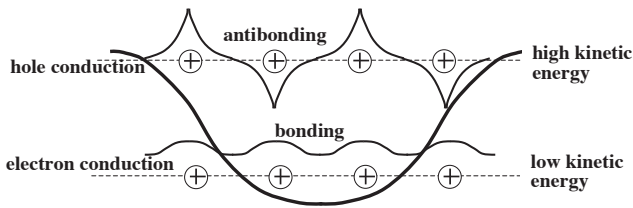}}
 \caption {When the Fermi level is near the top of the band, the kinetic energy of electrons at the Fermi level is high,  conduction
 occurs through hole carriers in the normal state, and the electronic charge density in the region between the ions is low.}
 \label{figure2}
 \end{figure}
 
If indeed superconductivity is driven by kinetic energy lowering, it will occur when the kinetic energy of carriers in the 
normal state is high. That will be the case when bands are almost full, as shown schematically in Fig. 4 where the conduction
in the normal state occurs through holes rather than electrons. We have discussed extensively in previous papers the
vast empirical evidence indicating that it is always $hole$ $carriers$ that drive superconductivity, and that materials without
hole carriers cannot be superconductors\cite{materials}.

   \begin{figure}
 \resizebox{8.5cm}{!}{\includegraphics[width=9cm]{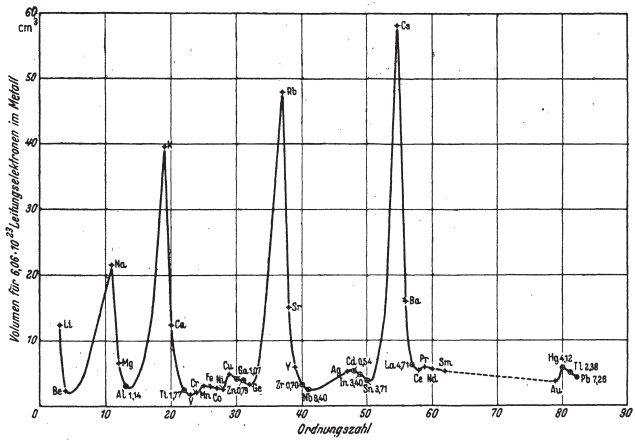}}
 \caption {Meissner-Schubert diagram (from ref. \cite{justi}). The ordinate gives the electron volume in 
 $cm^3$ for $6\times10^{23}$ conduction electrons. Some of the transition temperatures are given next to the elements.}
 \label{figure2}
 \end{figure}

Another way to put it:  if superconductivity involves an $expansion$ of the electronic wavefunction as discussed in Section II,
it should predominantly occur in systems where the electronic wavefunction is very $compressed$ in the normal state, so that
the drive to expand to lower the quantum kinetic energy is highest. It is interesting that precisely this criterion was
formulated by W. Meissner and G. Schubert in 1943\cite{ms}. They defined the quantity $V_E$, volume per valence electron, as
the difference between the atomic volume and the ionic volume, divided by the number of conduction electrons per atom,
and noted that superconductivity is associated with low values of $V_E$,
with the
smallest $V_E$ values corresponding to the highest critical temperatures.
Figure 5 shows the Meissner-Schubert diagram obtained from a paper by E. Justi\cite{justi}, where  $V_E$ is plotted versus atomic number. Justi remarks\cite{justi} that this
is a very marked correlation 
(``ausgepr\"{a}gte Regelm\"{a}ssigkeit'') that has ``evident truth content'' (``offenbare Wahrheitsgehalt''). It is also interesting to note
that in the same paper Justi remarks that there is no correlation between superconductivity and Debye temperatures:
``...finden wir weder eine spezielle Auszeichnung der
S-Leiter durch besondere $\theta_D$ Betr\"{a}ge, die vielmehr uber den gesamten vorkommenden Bereich von
$\theta_D=69^o$ bis $\theta_D=400^o$ streuen, noch insbesondere einen Zusammenhang zwischen $\theta_D$ and $T_c$''.\cite{justi2}

It would be interesting to extend the Meissner-Schubert diagram to many more superconducting elements and compounds discovered
since then.

 B. Matthias has  emphasized that superconductivity is very frequently associated with lattice instabilities\cite{matthias}. This observation follows
 naturally from the principles discussed here: first, carriers having high kinetic energy in the normal state gives rise to an
 unstable situation, and  to lower the kinetic energy  either the system will go superconducting  or the lattice will distort. 
Second, when the Fermi level is near the top of the band the electronic wavefunction is $antibonding$, with small
 charge density between the ions, in contrast to the $bonding$ wavefunction for electrons near the bottom of the band,
 as shown schematically in Fig. 4. The small electronic charge density between ions for antibonding electrons
 gives rise to repulsion between the ions and ``antibinds'' the lattice, thus leading to lattice instabilities. Thus, the fact that 
 many antibonding states are occupied by electrons (Fermi level near the top of the band) is associated with both
 lattice instabilities and superconductivity.
 
 In contrast, as is well known BCS theory does not particularly care about superconducting materials having low values
 of $V_E$ nor predominantly hole carriers in the normal state. It ascribes the prevalence of lattice instabilities near 
 superconductivity to a strong electron-phonon interaction. But it does not provide  simple criteria to predict which materials will
 have ``strong electron-phonon interaction'' and which will not; this is usually found out only after elaborate calculations that
 calculate $T_c$ after its value has been measured experimentally\cite{validity}. And BCS theory has nothing to say about the
 Meissner-Schubert correlation discussed above nor about  the necessity of hole carriers for
 superconductivity to occur, which many workers have pointed out in the past\cite{chapnik}.

\section{some experimental consequences}
We have discussed a variety of experimental consequences in previous papers. Here we focus on three:
\subsection{Plasmon dispersion relation}

The dispersion relation for longitudinal bulk plasmons in a normal metal is
\beq
\omega_q^2=\omega_p^2+\frac{3}{5} v_F^2q^2
\eeq
with $w_p^2=4\pi n e^2/n$, $n$ the electron density, and $v_F$ the Fermi velocity. Conventional BCS theory predicts that plasmons are
essentially unchanged in the superconducting state\cite{bcsplasmons,bcs2}. Instead, we predict\cite{electrodyn} for longitudinal charge oscillations of the
superfluid the dispersion relation
\beq
\omega_q^2=\omega_p^2+c^2q^2   
\eeq
which is  a significant change since $v_F<<c$ (we are assuming no change in the plasma frequency $\omega_p$ between normal and superconducting states for simplicity).

At finite temperatures the response should have a normal and a superfluid component. If $n_n$, $n_s$ are the normal and superfluid
densities at temperature $T$ we have 
\bmath
\beqn
\omega_q^2&=&\omega_p^2+[\frac{n_s}{n}c^2+\frac{n_n}{n}\frac{3}{5} v_F^2]q^2 \nonumber \\
&=& \omega_p^2+v_{eff}^2q^2
\eeqn
\beq
v_{eff}=\sqrt{\frac{n_s}{n}c^2+\frac{n_n}{n}\frac{3}{5} v_F^2}
\eeq
\emath
so that the slope of the plasmon dispersion relation $\omega_q$ should increase    from $\sqrt{3/5} v_F$ to $c$ as
$T$ is lowered from $T_c$ to $0$. The superfluid density at temperature $T$ is given by
\beq
n_s=n \frac{\lambda_L^2}{\lambda_L^2(T)}
\eeq
where $\lambda_L$ in the numerator is the zero-temperature London penetration depth. In a two-fluid model description one has
approximately $n_s=n(1-t^4)$, $n_n=nt^4$, with $t=T/T_c$.

To our knowledge the plasmon dispersion relation in superconductors has never been carefully studied experimentally, presumably because
no change is expected within BCS theory since the energies involved are much larger than the superconducting energy gap. In ref. \cite{plasmons},
N\"{u}cker et al reported measurements of EELS spectra in $Bi_2Sr_2CaCu_2O_8$ at room temperature and at the end of the paper stated
{\it ``we would like to mention that we have performed
similar measurements on excitations of valence
and core electrons at 30 K which is well below the superconducting
transitions temperature T, 83 K. Neither
the loss function nor the plasmon dispersion show a
significant difference between room temperature and 30
K.''} We believe the experiment should be repeated, since the theory discussed here  predicts a significant change in the
plasmon dispersion relation in that temperature range.
Our theory also predicts a significant change in the dispersion relation of surface plasmons  below $T_c$. 

\subsection{Screening and compressibility}

The dispersion relation Eq. (39) arises from the zeros of the longitudinal dielectric function of the superfluid which is very different from the Linhardt function of the
normal metal according to our theory\cite{electrodyn}. In the static limit this dielectric function   is \cite{electrodyn,comment}
\beq
\epsilon_s(q,\omega\rightarrow 0)=1+\frac{1}{\lambda_L^2 q^2}
\eeq
in contrast to the Linhardt-Thomas-Fermi form valid in the normal state as well as in the superconducting state within BCS theory\cite{bcsplasmons,bcs2}
\bmath
\beq
\epsilon_{TF}(q)=1+\frac{1}{\lambda^2_{TF}q^2}
\eeq
with
\beq
\frac{1}{\lambda_{TF}^2}=4\pi e^2 g(\epsilon_F)
\eeq
\emath
with $g(\epsilon_F)$ the density of states at the Fermi energy. These equations imply that static external electric fields are screened over distances
$\lambda_L$ and $\lambda_{TF}$ for the superconductor and the normal metal respectively.
For free electrons we have $g(\epsilon_F)=3n/2\epsilon_F$
so that
\beq
\frac{1}{\lambda_{TF}^2}=\frac{6\pi ne^2}{\epsilon_F}=\frac{1}{\lambda_L^2} \frac{3m_ec^2}{2\epsilon_F}
\eeq
assuming the density of superconducting electrons $n_s$ is the same as that of normal electrons. 
From the compressibility sum rule 
\beq
\epsilon(q\rightarrow 0,0)=1+\frac{4\pi e^2}{q^2}n^2\kappa
\eeq
with $\kappa^{-1}=-V\partial P/\partial V)_N$ it follows that the superconductor is much more rigid than the normal metal with respect to longitudinal
charge distortions:
\bmath
\beq
\kappa_s=\frac{1}{4\pi e^2 n_s^2 \lambda_L^2}=\frac{1}{n_sm_e c^2}  
\eeq
\beq
\kappa_n=\frac{1}{4\pi e^2 n^2 \lambda_{TF}^2}=\frac{g(\epsilon_F)}{n^2}=\frac{3}{2n \epsilon_F}
\eeq
\emath
(the latter valid for a free electron gas). Furthermore the superfluid will propagate longitudinal charge oscillations at the speed of light rather than
the Fermi velocity.

Thus, as the temperature is lowered below $T_c$, the electronic bulk modulus (inverse compressibility) should  increase rapidly as electrons condense into the superfluid phase.
This is not predicted by conventional BCS theory and should provide an explanation for the anomalous behavior of electronic sound propagation found 
by Avramenko et al\cite{sound} in 
metals cooled into the superconducting state.

\subsection{NQR}

Nuclear quadrupole resonance spectroscopy (NQR) measures the interaction between the electric quadrupole moment of a nucleus and the electric field gradient at the site.
Thus it may offer the possibility of detecting the internal electric fields in superconductors predicted by our theory.

In the superconducting state we predict an internal electric field that grows linearly with the distance to the center
as the surface is approached, reaching maximum value $E_m$ given by
Eq. (17) at distance $\lambda_L$ from the surface and decreasing to zero approximately linearly as the surface is reached. Thus the electric field gradient
created by this electric field within distance $\lambda_L$ of the surface is
\beq \label{eq:nqr}
\frac{\partial \vec{E}}    {\partial \hat{n}} \sim \frac{E_m}{\lambda_L}
\eeq
For example, for $Nb$, $Pb$ and $In$ we have respectively $E_m=308,400 V/cm$, $\lambda_L=400 \AA$, $E_m=240,900 V/cm$, $\lambda_L=390 \AA$ and
$E_m=87,900 V/cm$, $\lambda_L=640 \AA$.

The electric field gradient Eq. (\ref{eq:nqr})  is   too small to be detected directly. For example, the electric field gradient in $In$ metal measured by NQR
is of order $10^{18} V/cm^2$\cite{nqr1}, several orders of magnitude larger than what Eq. (\ref{eq:nqr}) predicts. However the electric field itself will shift ionic positions and modify the electronic charge distribution around the nuclei which 
will change the electric field gradient at the nuclear site and this effect should be large enough to be observable. For example, application of an electric
field of magnitude $17,100 V/cm$ to $KClO_3$ was found to result in a substantial shift and change of lineshape of the $Cl^{35}$ quadrupole resonance\cite{nqr2}.

In 1961 Simmons and Slichter\cite{nqr3} reported a marked shift in the nuclear quadrupole resonance frequency of $In$ below the superconducting transition temperature,
of approximately $2\%$ downwards, as well as a change in lineshape from symmetric to asymmetric. They remarked that 
{\it ``volume changes associated with the superconducting phase transition are not, by several orders of magnitude, large enough to account for this
large shift''} and they concluded that {\it ``The explanation of the large shift remains an open question at present.''} It is still an open question today, and
the experiment has never been repeated.

There have been remarkably few other measurements of NQR resonance in zero magnetic field in the superconducting state. Hammond and Knight\cite{nqr4} reported a very small
frequency shift in the superconducting state of $Ga$. In $YBCO$, a steep decrease in the frequency as $T_c$ was approached from above was measured, and
an increase below $T_c$\cite{nqr5}. That reference also gives a complete list of NQR measurements in the normal and superconducting states of various materials
over the years. The list is remarkably short.

We suggest that a systematic study of NQR frequency shifts and lineshape changes between the normal and superconducting state of various materials, both
those categorized as `conventional' and `unconventional', would be of great interest. No such effects are expected within the conventional theory of superconducitivity nor
have such effects been predicted within other unconventional theories. Measurement of such effects would provide direct evidence for the development of 
an electric field in the interior of superconductors as they enter the superconducting state, as   predicted by our theory.

\section{superfluid pressure and relation with  $^4He$}

   \begin{figure}
 \resizebox{8.5cm}{!}{\includegraphics[width=9cm]{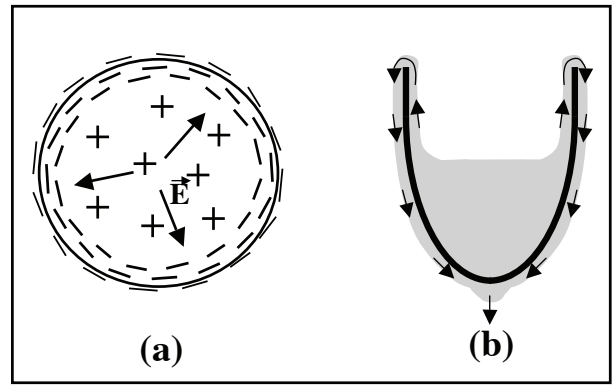}}
 \caption{Superfluid pressure causes (a) electrons in superconductors to be expelled from the interior to the surface and beyond, and
(b)  $^4 He$ to climb the lateral surfaces of a container and escape to the exterior (``Onnes effect'').}
 \label{figure2}
 \end{figure}
 
The essential common aspect of the physics of superconductors and superfluid $^4He$, unrecognized in the currently accepted theories of both systems,
appears to be  that {\it superfluids exert quantum pressure}  that is larger than in the normal state. This is in contrast to the situation in 
conventional Bose condensation where the condensate exerts no pressure.

In other words, as the system goes superfluid or superconducting and long range coherence sets in, the wavefunction of the mobile carriers exerts additional outward pressure
and expands its spatial extent, driven by lowering of
quantum kinetic energy.

For superconductors, this quantum pressure manifests itself in the negative charge expulsion (not yet experimentally detected) that we predict exists in all superconductors 
(Fig. 6(a)) and
associated with it in the ``Meissner pressure''\cite{londonpressure} that expels the magnetic field
through orbit expansion, opposing the ``Maxwell pressure'' that wants to keep the magnetic field inside.
As discussed in Sect. II, Meissner pressure is proposed to originate in quantum pressure which in turn is proposed to
originate in the fact that a rotating body with fixed angular momentum lowers its
kinetic energy by expanding its orbit. In is interesting to note that in the early development of electromagnetism, Maxwell explained the 
``Maxwell pressure'' exerted by magnetic fields in direction perpendicular to the field also as originating in rotational motion (of
``molecular vortices'')  with axis along the magnetic 
field direction\cite{maxwellpressure}. 

Another manifestation of the superfluid pressure for the case of superconductors is that tunneling currents in NIS tunnel junctions are larger for
negatively biased samples (Fig. 7 left panel)\cite{tunnasym,tunn2}, reflecting the tendency of superconductors to expel electrons. Yet another manifestation is the proximity effect,
where the superconducting wavefunction extends from the superconducting into the neighboring normal region.

For superfluid $^4He$\cite{londonbooks}, the superfluid pressure manifests itself vividly in the ``fountain effect'': when the superfluid concentration is depleted in one region of a 
container by heating, the superfluid from another region will spurt into the depleted region with great force, driven  by this pressure (Fig. 7 right panel). It also manifests itself
(according to our interpretation)\cite{helium}
in the ``Onnes effect'', the fact that the superfluid will climb up the walls of a container, defying gravity (Fig. 6(b)). It manifests itself in the anomalous negative thermal
expansion of superfluid $^4He$ below the $\lambda$ point.  It manifests itself in the $inverted$  $\lambda-$shape of the specific heat curve in $^4He$ (that gives the name to the
$\lambda-$ transition) which is direct evidence for quantum kinetic energy lowering in the transition to superfluidity\cite{lambda}.
All these phenomena we argue are manifestation of {\it superfluid pressure} originating in 
quantum pressure i.e. kinetic energy lowering. We have proposed that, just as in superconductors, in superfluid $^4He$ this pressure originates in {\it rotational zero point motion}\cite{helium}.

   \begin{figure}
 \resizebox{8.5cm}{!}{\includegraphics[width=9cm]{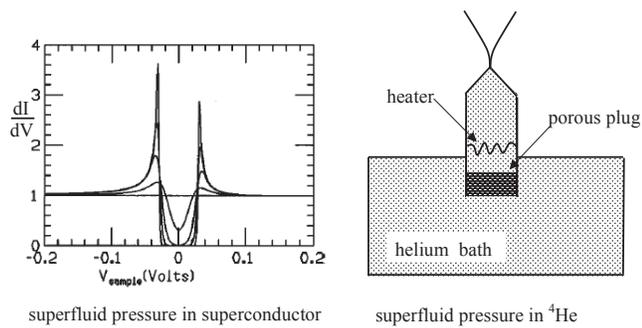}}
 \caption{Illustration of manifestations of superfluid pressure. In superconductors (left), it gives rise to asymmetric tunneling characteristics, frequently observed in high 
 $T_c$ materials, reflecting (according to our theory) the pressure pushing out negative electrons from the interior of superconductors. In superfluid $^4He$ (right) it gives rise to the fountain 
 effect, reflecting the superfluid pressure that drives the superfluid  flow into the hotter region that has lower superfluid density. }
 \label{figure2}
 \end{figure}
 
Furthermore, the enhanced charge rigidity that we predict in the superconducting state is clearly associated with this quantum pressure and 
has its counterpart in $^4He$ in the enhanced bulk
 modulus (inverse compressibility) observed  below the $\lambda-$ point\cite{compresshe}. 

For superconductors we have shown that electron-hole asymmetry gives rise to $positive$ thermoelectric power for NIS tunnel junctions\cite{thermopower}. 
Namely, if the superconductor is at lower temperature than the normal electrode, electrons will flow $from$ the superconductor $to$ the normal electrode,
exactly the opposite to what would happen in the normal state. This is entirely analogous in the case of $^4He$: above $T_\lambda$, 
the normal fluid will flow from the hotter to the colder region in a container. Instead, below $T_\lambda$ the dominant flow is from the colder region to the
hotter region (as in the fountain effect). This is again clear illustration of the identical effect of quantum pressure for superconductors and 
superfluids  that we propose is responsible for
these phenomena.

In superfluid $^4He$, a backflow of normal fluid flow occurs whenever the constraints of the system allow it when there is superfluid flow  from a colder to a hotter region
of the container. 
Similarly we may expect  such backflow to occur in superconductors: when electrons in the interior condense into the superfluid state and are expelled towards the surface
 in the transition to
superconductivity, the outward charge flow should  be partially compensated  by an $inflow$ of normal electrons. In the presence of a magnetic field,
the inflowing electrons will be deflected by the Lorentz force in opposite direction to the outflowing electrons and will transmit   that angular momentum
to the ionic lattice by scattering. This should play an important role in explaining the puzzle of how angular momentum is conserved when 
a metal goes superconducting in the presence of a magnetic field and the Meissner current that carries angular momentum develops\cite{missing}.

In summary, we have seen that several different phenomena in superfluid $^4He$ and superconductors can be explained if the superfluids exert quantum pressure
due to quantum zero point motion.
In addition, the frictionless flow of electric current in superconductors and the nonviscous flow of superfluid $^4He$ and
  their respective critical velocities can be explained in a unified way 
as originating in quantum zero point diffusion\cite{mendelflow,helium}.
Of course the fact that quantum zero point motion plays an important role in liquid $He$ is generally recognized. For example, everybody agrees that this is the reason why
$He$ remains liquid under its own vapor pressure down to zero temperature. However the fact that superfluid
$^4He$ exerts quantum pressure that is larger than that of the normal fluid is not generally recognized, despite the clear experimental evidence for it.
A notable exception can be found in the writings of K. Mendelssohn who consistently emphasized this key aspect of superfluid $^4He$, for example when he
writes\cite{mendpressure}: {\it ``A question of particular interest is whether or not the superfluid phase contributes to the pressure. According to the Bose-Einstein model
(F. London, 1939) this contribution is zero, but it appears to us from the experimental results that such a zero-point pressure must exist and that it
is of fundamental importance for the explanation of the transport phenomena.''}

\section{discussion}
 
  We argue that the Meissner effect is an unresolved puzzle within the conventional theory of superconductivity. 
How can superconductors
  governed by conventional BCS-London theory  expel a magnetic field when cooled from the normal into the superconducting state?
The magnetic Lorentz force
\beq \label{eq:last}
\vec{F}=\frac{e}{c}\vec{v}\times\vec{B}=\frac{e}{c} B(v_r\hat{\theta}-v_\theta\hat{r})\equiv F_\theta \hat{\theta}+F_r\hat{r}
\eeq
will not give rise to an azimuthal force $F_\theta$ that will set the Meissner current into motion $unless$ there is a $radial$ velocity $v_r$, i.e. a net
outflow of charge:  in Eq. (\ref{eq:last}), $F_\theta=0$ if $v_r=0$.
BCS-London theory does not describe radial flow of charge, hence $v_r=0$. Unless and until proponents of BCS-London theory explain which force in nature will propel the charge
near the surface to move in the
azimuthal direction   and overcome  the Faraday counter-emf to generate the
Meissner current, the Meissner effect will remain unexplained within the conventional theory.
Since all superconductors exhibit the Meissner effect, we must conclude that {\it the conventional theory in its present form does not apply to any real superconductor}.

  Instead, we have proposed that the Meissner effect in all superconductors is explained by orbit expansion driven by quantum pressure, that 
  gives rise to outflow of negative charge, an outward pointing  electric field in the interior of superconductors, an excess negative charge near the surface, enhanced
  charge rigidity, rotational zero point motion in $2\lambda_L$ orbits reflecting electronic Zitterbewegung, and
  a macroscopic spin current near the surface. This non-conventional physics should exist in all superconductors that exhibit the Meissner effect. That is, 
  in all new and old superconductors, `conventional' and `unconventional'.
  
  Concerning materials, this physics indicates that superconductivity should be particularly favored in materials 
  that have a lot of negative charge, namely almost filled bands (hole conduction in the normal state) and negatively charged anions, as well as highly compressed
  electronic wavefunctions giving rise to high kinetic energy and quantum pressure, since all of these factors will contribute to the tendency to expel negative
  charge. Materials evidence in favor of this\cite{materials} is the high $T_c$ observed in the 
  cuprates, pnictides and $MgB_2$, all possessing negative anions, that superconductors overwhelmingly display dominant $hole$ conduction in the normal state
  (positive Hall coefficient), that they show a particularly small volume per electron (Meissner-Schubert diagram) and that they are often close 
  to lattice instabilities,  indicating high occupation of $antibonding$ states.
  
  Furthermore, we have argued that 
   this view of superconductivity has close connection to and sheds new light into the physics of superfluidity in $^4He$. Both  superconductivity and superfluidity 
  are proposed to be {\it kinetic energy driven}, due to the fact that superfluids exert quantum pressure.  This suggests a common explanation for many observed  
  phenomena in superfluids and superconductors that would be unrelated otherwise, such as the superfluid fountain effect, the tunneling asymmetry in superconductors, 
  the thermomechanical effect in $^4He$, the predicted positive thermoelectric power in superconducting tunnel junctions,
the Onnes effect in superfluid $He$ films, the Meissner effect in superconductors, the transfer of optical spectral weight from high to low frequencies in
  superconductors, and the negative thermal expansion of superfluid $He$. The non-relativistic virial theorem, which says that kinetic energy should be raised
  when the total energy is lowered if the dominant interactions in the system are Coulomb, is profoundly misleading for both superconductors and superfluids.

The physics of superconductivity as driven by kinetic energy lowering and exhibiting rotational zero point motion 
   has led us to conclude that {\it the fundamental origin of quantum pressure  
 in nature   is  rotational zero-point motion}, and in particular is responsible for the stability of matter and the Pauli exclusion principle\cite{double,kinetic}. 
 In contrast,   in the conventional understanding of quantum mechanics there is no rotational zero point motion, and
  the origin of quantum pressure and the stability of matter is attributed to Heisenberg's uncertainty principle. However, an alternative 
  interpretation of quantum mechanics connecting Heisenberg's 
  uncertainty principle to an intrinsic rotational motion of the electron due to its spin was already proposed in pioneering work 
  by D. Hestenes in 1979\cite{hestenes2}. The idea that rotational motion 
 gives rise to pressure is of course very old: it was at the core of the early description of magnetic fields by Maxwell 
  (molecular vortices)\cite{maxwellpressure}  as well as in the theory of
  ``vortex atoms'' by Lord Kelvin\cite{vortexatoms}. 
Our  suggestion derived from both the physics of superconductors (fermions) and superfluid $^4He$ (bosons),  that rotational zero point motion   is  at the root of
  quantum pressure quite generally, suggests   that rotational zero point motion may be a fundamental element
  of the fabric of space-time itself, and it is natural to speculate that it may also be at the root of other phenomena such as the enigmatic dark energy that pervades the
  universe\cite{darkenergy}.

\acknowledgements{The author is grateful to R. Orbach for suggesting  NQR as a probe of the predicted electric field in superconductors},
and to F. Marsiglio, A.S. Alexandrov, A. Bussmann-Holder, H. Keller , A.J. Leggett and J. Fink for stimulating comments and discussion.


\begin{references}
\bibitem{justi2} Translation: ``...we find neither a special distinction of superconductors as having particular values of $\theta_D$ (Debye temperature),
which in fact spread over the entire range of values from $\theta_D=69^o$ to
$\theta_D=400^o$, nor in particular any relationship between $\theta_D$ and $T_c$ (superconducting critical 
temperature''.
\end{references}
\end{document}